# Searching for Patterns among Squares Modulo *p*

Roger Bilisoly[1]

[1]Department of Mathematical Sciences, Central Connecticut State University,
1615 Stanley St., New Britain, CT 06050-4010

**Abstract**

Although squaring integers is deterministic, squares modulo a prime, *p*, appear to be random. First, because they are all generated by the multiplicative linear congruential equation, $x_{i+1} = g^2 x_i \pmod{p}$, where $x_0 = 1$ and *g* is any primitive root of *p*, a pseudorandom number heuristic suggests that they are, in fact, unpredictable. Moreover, one type of cryptography makes use of discrete algorithms, which depends on the difficulty of solving $a = g^n$ for *n* given *a* and *g*. This suggests that the squares, which are exactly the even powers of *g*, are hard to identify. On the other hand, the Legendre symbol, (*a*/*p*), which equals 1 if *a* is a square (mod *p*) and -1 otherwise, has proven patterns. For example, (*ab*/*p*) = (*a*/*p*)(*b*/*p*) holds true, and this shows that squares modulo *p* have some structure. This paper considers the randomness of the following sequence: (1/*p*), (2/*p*), …, ((*p*–1)/*p*). Because it consists of binary data, the runs test is applied, which suggests that the number of runs is exactly (*p*–1)/2. This turns out to be a theorem proved by Aladov in 1896 that is not widely known. Consequently, this is an example of a number theory fact that is revealed naturally in a statistical setting, but one that has rarely been noted by mathematicians.

**Key Words:** Pseudorandom numbers, Number theory, Legendre symbol, Runs test

## 1. Introduction

Mathematicians prove theorems, but this requires the ability to generate plausible claims to be proven. One way to do this is by using heuristics. For example, by assuming prime numbers are randomly distributed, one can estimate the probability of *n* being prime is about 1/log(*n*) by the prime number theorem (PNT). However, theorems about primes are often difficult to prove. For instance, again using the PNT, Hardy and Littlewood conjectured that the expected number of prime twins less than *n* should be proportional to $n/(\log(n))^2$, but no one has proven that there are even an infinite number of prime twins.

The goal of this paper is to use heuristics based on statistical ideas from areas such as pseudorandom number generation and nonparametric rank statistics to generate potential theorems. Because statements about primes can be hard to prove, this paper looks at squared integers modulo *p*, a prime. That is, we consider the remainders of squares divided by *p*. Number theory and abstract algebra tell us much about this situation, and the results found below have all been proven. This suggests that the statistical analysis of data arising from studying mathematical structures could lead to new results that are provable.

## 2. Squares Modulo *p*

Subsection 2.1, using a heuristic, gives a plausible argument that squares modulo *p* are randomly distributed. However, the subsequent subsection describes how quadratic reciprocity, a theorem from number theory, proves that these squares must follow certain patterns. Because these are in conflict, we perform hypothesis tests of empirical data to see if any deviations from randomness are detectable.

### 2.1 Congruential Pseudorandom Numbers

After the computer was invented, one of its first tasks was generating random numbers for Monte Carlo simulations. According to Section 2.1 of Ripley (1987), in 1951 D. H. Lehmer published his work on the multiplicative congruential generator, which is given by Equation (1).

$$x_{i+1} = a\, x_i \,(\mathrm{mod}\ m), \qquad (1)$$

where *a* and *m* are integers, and set $x_0 = 1$. However, it turns out that an equation of this form also generates all the squares modulo *p*, which suggests that these may display randomness, too.

It turns out that not all choices of *a* and *m* produce good results. For example, if *a* is quite small compared to *m*, then two consecutive values of *x* are never close together. Unfortunately, subtler problems can occur. For instance, setting $m = 2^{13} - 1 = 8191$, $a = 1904$, then Equation (1) produces 1, 1904, 4794, 3002, 6681, 1, …. That is, it only generates five distinct values and then starts to repeat. This suggests a heuristic: choose values of *a* and *m* so that the number of distinct values is as large as possible.

The maximum possible number of distinct values is $m - 1$. Call the number of distinct values the *period*, then Theorem 2.3 of Ripley (1987) states which periods are possible. At the end of the 18th century Gauss proved that for every prime, there is at least one value of *a* that has the maximal period: see Articles 52-54 of Gauss (1966). Such values are called *primitive roots*.

**Theorem 2.3 of Ripley (1987).** *A multiplicative generator has period $m - 1$ only if m is prime. Then the period divides $m - 1$, and is $m - 1$ if and only if a is a primitive root.*

If *g* be a primitive root of *p*, an odd prime, then Equation (2) with $x_0 = 1$ generates all the squares modulo *p*. Because these have period $(p - 1)/2$, the second longest possible value, the above heuristic suggests this sequence should be fairly random.

$$x_i = g^2\, x_{i-1}\,(\mathrm{mod}\ p), \qquad (2)$$

Finally, even with large periods, Equation (1) can give poor results, and this is still true even if a constant term is added: see the discussion in Sections 2.2 and 2.4 of Ripley (1987). Nonetheless, although better methods are employed today (for example, the yarrow algorithm or methods used for cryptography), this approach was successfully used to generate random numbers for many years.

## 2.2 Quadratic Reciprocity

The preceding subsection gives a heuristic reason to believe squares modulo $p$ should be random, but quadratic reciprocity from number theory shows that there are some constraints, too. For $p$ an odd prime and $a$ not a multiple of $p$, we define the Legendre symbol, $(a/p)$, to be 1 exactly when $a$ is a square modulo $p$, and -1 otherwise. The following theorems are proven in Chapter 5 of Ireland and Rosen (1990).

**Theorem.** *For an odd prime p that does not divide a nor b, (ab/p) = (a/p)(b/p) holds.*

**Theorem (Quadratic Reciprocity).** *For odd primes p and q, (p/q) = -(q/p) exactly when both p and q are congruent to 3 (mod 4), otherwise (p/q) = (q/p).*

Using the above results, the following equations can be proved, which show that squares are quite predictable.

$$(-1/p) = 1 \text{ iff } p \equiv 1 \pmod{4}$$
$$(2/p) = 1 \text{ iff } p \equiv \pm 1 \pmod{8}$$
$$(3/p) = 1 \text{ iff } p \equiv \pm 1 \pmod{12} \quad (3)$$
$$(5/p) = 1 \text{ iff } p \equiv \pm 1 \pmod{5}$$
$$(6/p) = 1 \text{ iff } p \equiv \pm 1, \pm 5 \pmod{24}$$
$$\ldots$$

For instance, since $p = 8191 \equiv -1 \pmod 8$, 2 is a square. However, finding the "square root" requires solving $g^l = 2$ for $l$, where $g$ is a primitive root of $p$, which is an example of discrete logarithms, a method used in cryptography because it is computationally difficult for large $p$. Of course, for small primes, brute force can be used: $128^2 \equiv 2 \pmod{8191}$.

By this and the last subsection, squares modulo an odd prime cannot be entirely random, but there is reason to believe there is some unpredictability. This suggests applying statistical tests of randomness, which is done in the next section.

## 3. Tests of Randomness for Squares Modulo *p*

Many tests have been developed for randomness. For example, the National Institute of Standards and Technology (NIST) has developed a test suite as described in Rukhin et al. (2010). This section uses two different tests, both of which will detect patterns in squares modulo *p*.

### 3.1 Metrics on the Symmetric Group, $S_{p-1}$

The first approach is to look at powers of a primitive root, which by definition produce a permutation of all the numbers from 1 through $p - 1$. In general, there are $\phi(p - 1)$ primitive roots, where $\phi$ is the Euler totient function, so testing randomness in this context can be viewed as picking $\phi(p - 1)$ random $(p - 1)$-cycles from $S_{p-1}$. In this paper, these will always be written with $1 = g^0$ first. As a simple example, consider $p = 11$, which has the four primitive roots 2, 6, 7, 8. Hence we have four 10-cycles: (1, 2, 4, 8, 5, 10, 9, 7, 3, 6), (1, 6, 3, 7, 9, 10, 5, 8, 4, 2), (1, 7, 5, 2, 3, 10, 4, 6, 9, 8), and (1, 8, 9, 6, 4, 10, 3, 2, 5, 7). Notice that the squares are the even powers, and each of these cycles gives the same set of squares: {1, 3, 4, 5, 9}. Recall that $1 = g^0$, so the even powers are in the odd positions of each cycle.

To decide how random these 10-cycles are, we need a metric to measure the distance of these from the extreme case, (1, 2, 3, 4, 5, 6, 7, 8, 9, 10). It turns out that the symmetric group has many metrics, which have been well studied. Moreover, close links between these metrics and nonparametric statistical tests are known. For example, see Critchlow (1986) and Chapter 6 of Diaconis (1988).

Here we consider only one metric, the number of inversions in a permutation, which is the number of times a larger integer is to the left of a smaller one. For instance, there are 15 of them in (1, 2, 4, 8, 5, 10, 9, 7, 3, 6) because of the following: 10 is to the left of 9, 7, 3, 6; while 9 is to the left of 7, 3, 6; and so forth until we stop with 4 being to the left of 3.

For a specific example consider $p = 29$. There are 12 primitive roots, {2, 3, 8, 10, 11, 14, 15, 18, 19, 21, 26, 27}, and each of these generates a 28-cycle. For instance, 2 produces (1, 2, 4, 8, 16, 3, 6, 12, 24, 19, 9, 18, 7, 14, 28, 27, 25, 21, 13, 26, 23, 17, 5, 10, 20, 11, 22, 15). Page 117 of Diaconis (1988) proves that the mean is $(p - 2)(p - 3)/4$, while the variance is $(p - 2)(p - 3)(2p + 1)/72$ for the number of inversions. Note that these expressions have been modified to take into account that 1 is always first.

The actual number of inversions are 129, 159, 168, 192, 183, 171, 222, 194, 205, 157, 146, 180, respectively. The mean of these is 175.5, which is exactly the theoretical mean, $(p - 2)(p - 3)/4 = 27*26/4 = 175.5$. Second, the sample standard deviation is 26.02, which is bigger than the theoretical value of 23.98. Are the equality of means an accident? Is 26.02 significantly bigger than 23.98? To answer these questions, 10,000 random 28-cycles were generated, and Figure 1 gives the histogram of the number of inversions.

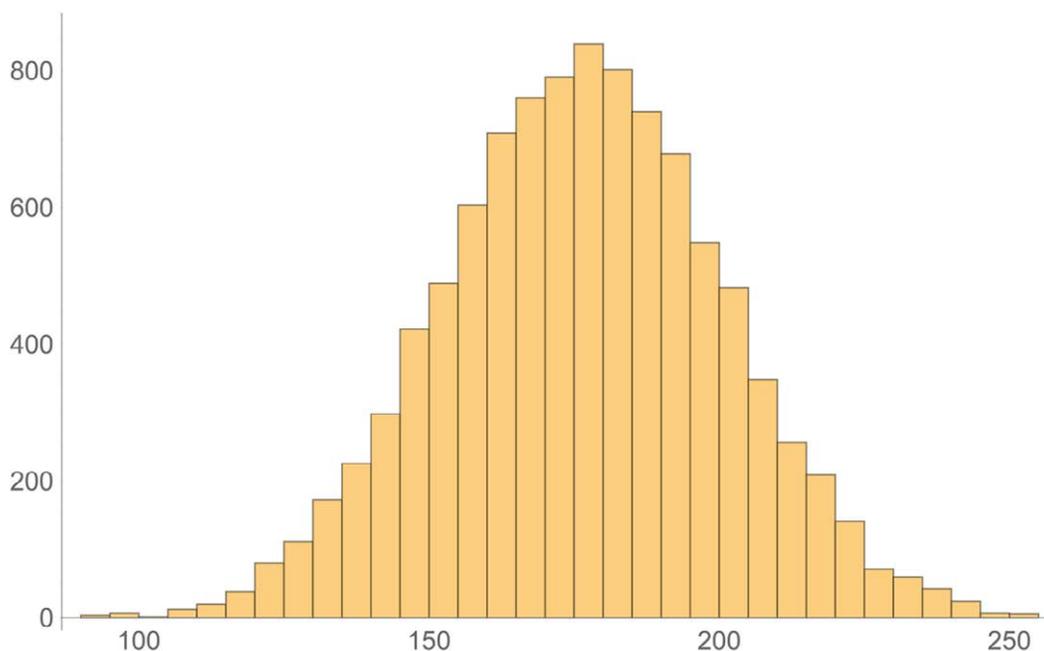

**Figure 1:** For $p = 29$, the number of inversions of 10,000 randomly generated cycles from $S_{p-1} = S_{28}$ that start with 1.

The mean of the 10,000 inversions is 175.86, which is close but not equal to 175.5. This suggests that getting an exact match with a sample size of 12 is suspicious, and it turns out that it is easy to prove that equality must happen. There is a natural pairing of $(p - 1)$-cycles because primitive roots come in pairs: $g$ is one if and only if $g^{-1}$ is one, too. For example, 2 and 15 are inverses, and these produce the two cycles given in Equation (4). Besides the initial 1, each cycle is in the reverse order of the other, hence the number of inversions of both of them add to the total number of inversions, which is given by $(p - 2)(p - 3)/2$. Consequently, averaging these two numbers gives $(p - 2)(p - 3)/4$, which is the theoretical mean.

$$(1,2,4,8,16,3,6,12,24,19,9,18,7,14,28,27,25,21,13,26,23,17,5,10,20,11,22,15) \quad (4)$$
$$(1,15,22,11,20,10,5,17,23,26,13,21,25,27,28,14,7,18,9,19,24,12,6,3,16,8,4,2)$$

Finally, the sample standard deviation of the 10,000 inversions is 24.11, which is quite close to the theoretical value, 23.98. This suggests that standard deviation of the 12 28-cycles generated by the primitive roots, 26.02, is significantly bigger than 23.98. Taken together, this is evidence that the squares modulo 29 are somewhat but not maximally random. This conclusion is reinforced by the results of the next subsection.

### 3.2 Runs Test Applied to the Legendre Symbol

In this subsection we test the randomness of the sequence $(1/p), (2/p), (3/p), \ldots, (p - 1/p)$. Because the Legendre symbol only has two values, 1 and -1, this is an example of binary data, which suggests using the runs test. However, some deviations from randomness are immediate. For instance, $(1/p) = (4/p) = (9/p) = \ldots = 1$ for all odd primes.

Because the squares modulo $p$ are exactly the even powers of any primitive root, $g$, there are precisely $(p - 1)/2$ squares and non-squares, hence we consider sequences of half 1s and half -1s. Under the null hypothesis of complete randomness, every ordering is equally likely, and Figure 2 gives an example of the histogram of the number of runs for 10,000 random permutations for $p = 97$, which looks approximately normal.

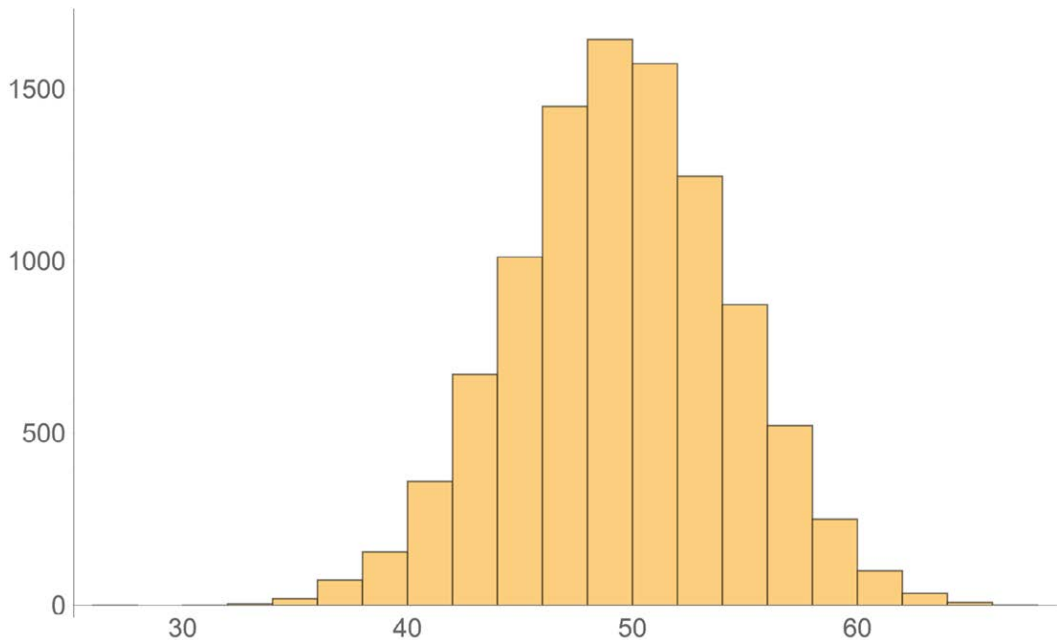

**Figure 2:** For $p = 97$, the number of runs for 10,000 random permutations of $(p - 1)/2 = 48$ 1s and -1s.

Because the minimum number of runs is 2, and the maximum is $(p - 1)$, Figure 2 suggests that the mean result might be at the halfway point, $(p + 1)/2$, and there is certainly variability. Now we compare this to the data in Figure 3 for the first seven odd primes. Surprisingly, in every case there are exactly $(p + 1)/2$ runs, and to check this, Figure 4 plots the number of runs versus $p$ for the first 200 odd primes, and the pattern still holds without exception.

$$p=3, (1, -1), \text{\# runs} = 2$$
$$p=5, (1, -1, -1, 1), \text{\# runs}=3$$
$$p=7, (1, 1, -1, 1, -1, -1), \text{\# runs}=4$$
$$p=11, (1, -1, 1, 1, 1, -1, -1, -1, 1, -1), \text{\# runs}=6$$
$$p=13, (1, -1, 1, 1, -1, -1, -1, -1, 1, 1, -1, 1), \text{\# runs}=7$$
$$p=17, (1, 1, -1, 1, -1, -1, -1, 1, 1, -1, -1, -1, 1, -1, 1, 1), \text{\# runs}=9$$
$$p=19, (1, -1, -1, 1, 1, 1, 1, -1, 1, -1, 1, -1, -1, -1, -1, 1, 1, -1), \text{\# runs}=10$$

**Figure 3:** The number of runs for the sequence $(1/p), (2/p), (3/p), \ldots, (p - 1/p)$ for the first seven odd primes.

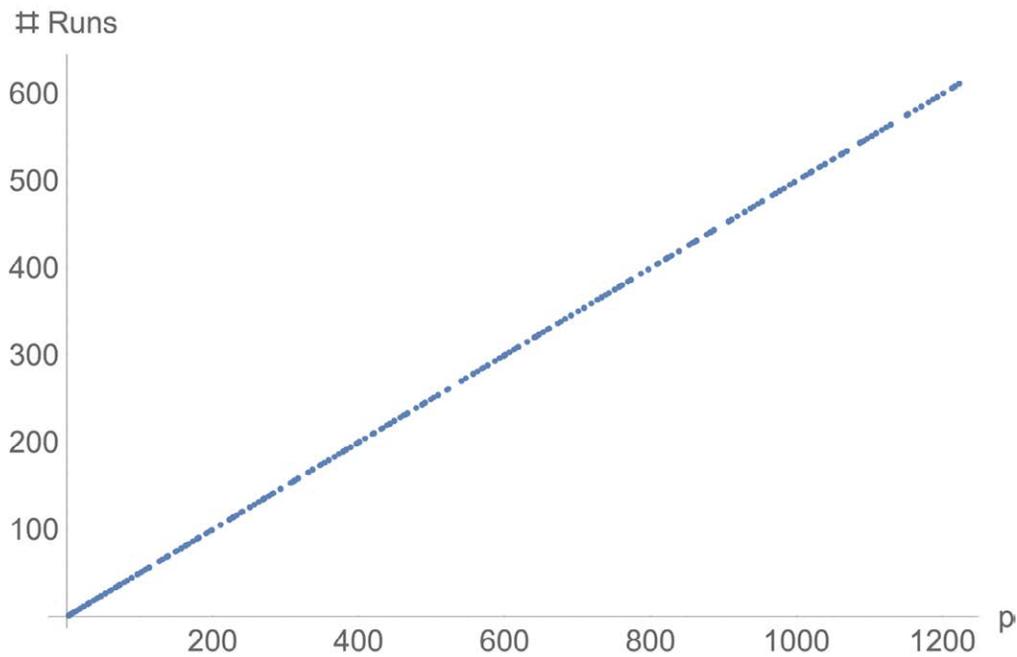

**Figure 4:** Plot of the number of runs versus $p$, which forms a straight line.

An initial search of the number theory literature turned up nothing, so I contacted Keith Conrad at the University of Connecticut, who is an expert in number theory (Conrad (2013)). He was unaware of this result, but was able to prove it within a day. Eventually a proof in the literature was found, and the theorem below is quoted in Peralta (1992) where it is attributed to Aladov, who published a proof in Russian in 1896.

Some notation is needed. Take the sequence $(1/p), (2/p), (3/p), \ldots, (p - 1/p)$ and break it into overlapping pairs. For example, for $p = 7$, $(1,1,-1,1,-1,-1)$ becomes $(1,1)$, $(1,-1)$, $(-1,1)$, $(1,-1)$, $(-1,-1)$. Let $n_{++}$ be the number of $(1,1)$s, $n_{+-}$ the number of $(1,-1)$s, $n_{-+}$ the number of $(-1,1)$s, and $n_{--}$ the number of $(-1,-1)$s. Then the following is true.

**Theorem (Aladov).** *For p ≡ 1 (mod 4), $n_{+-} = n_{-+} = (p-1)/4$, $n_{--} = (p-1)/4$, $n_{++} = (p-5)/4$. For p ≡ 3 (mod 4), $n_{++} = n_{--} = (p-3)/4$, $n_{+-} = (p+1)/4$, $n_{-+} = (p-3)/4$.*

Because the number of runs is the number of sign changes plus 1, this theorem implies the following.

**Corollary.** *For all odd primes p, the number of runs is $n_{+-} + n_{-+} + 1 = (p + 1)/2$.*

### 3.3 Conclusion

Given the increasing power of computers coupled with symbolic mathematics packages such as Sage, there has been a growing interesting in analyzing empirical data to formulate conjectures. For instance, the journal *Experimental Mathematics* is devoted to this approach. This paper gives two examples that show statistical thinking can generate fruitful mathematical leads that led to theorems.

### Acknowledgements

Thanks to Professor Keith Conrad, University of Connecticut, for taking time to answer my emails and for working out a proof of the above Corollary.